\documentclass[sigconf,nonacm]{acmart}

\renewcommand\footnotetextcopyrightpermission[1]{}
\usepackage{xspace}
\usepackage{xcolor}
\usepackage{url}
\usepackage{listings}

\usepackage{todonotes}

\usepackage{makecell}

\newcommand{\no}{\(x\)} 

\newcommand{\emfularc}{\textsc{EMF\-u\-lar-\-Core}\xspace}
\newcommand{\emfhelpers}{\textsc{EMF\-u\-lar-\-Tool}\xspace}
\newcommand{\emfsvg}{\textsc{EMF\-u\-lar-\-Di\-a\-gram}\xspace}
\newcommand{\emfintegration}{\textsc{EMF\-u\-lar-\-In\-te\-gra\-tion}\xspace}
\newcommand{\emfgenerator}{\textsc{EMF\-u\-lar-\-Ge\-ne\-ra\-tor}\xspace}
\newcommand{\emfularp}{\textsc{EMF\-u\-lar}\xspace}

\newcounter{rq}
\renewcommand{\therq}{RQ\arabic{rq}}
\newcommand{\researchquestion}[2]{%
  \refstepcounter{rq}%
  \item[\therq]\label{#1}\textbf{#2}%
}

\definecolor{lightgray}{rgb}{.9,.9,.9}
\definecolor{darkgray}{rgb}{.4,.4,.4}
\definecolor{purple}{rgb}{0.65, 0.12, 0.82}

\lstdefinelanguage{JavaScript}{
  keywords={typeof, new, true, false, catch, function, return, null, catch, switch, var, if, in, while, do, else, case, break},
  keywordstyle=\color{blue}\bfseries,
  ndkeywords={class, export, boolean, throw, implements, import, this, protected},
  ndkeywordstyle=\color{darkgray}\bfseries,
  identifierstyle=\color{black},
  sensitive=false,
  comment=[l]{//},
  morecomment=[s]{/*}{*/},
  commentstyle=\color{purple}\ttfamily,
  stringstyle=\color{red}\ttfamily,
  morestring=[b]',
  morestring=[b]"
}

\AtBeginDocument{%
  }
    
\setcopyright{acmlicensed}
\copyrightyear{2018}
\acmYear{2018}
\acmDOI{XXXXXXX.XXXXXXX}
\acmConference[Conference acronym 'XX]{Make sure to enter the correct
  conference title from your rights confirmation email}{June 03--05,
  2018}{Woodstock, NY}
\acmISBN{978-1-4503-XXXX-X/2018/06}

\setcopyright{none}
\renewcommand\footnotetextcopyrightpermission[1]{}

\begin{document}

\title{Web‑Native Graphical EMF Model Editors}

\author{Susanne G\"obel}
\email{goebel@uni-koblenz.de}
\orcid{0009-0002-6865-7167}
\affiliation{%
  \institution{SoftLang Team, Faculty of CS, University of Koblenz}
  \city{Koblenz}
  \country{Germany}
}
\author{Ralf L\"ammel}
\email{laemmel@uni-koblenz.de}
\orcid{0000-0001-9946-4363}
\affiliation{%
  \institution{SoftLang Team, Faculty of CS, University of Koblenz}
  \city{Koblenz}
  \country{Germany}
}

\renewcommand{\shortauthors}{G\"obel and L\"ammel}

\begin{abstract}
Graphical model editing is shifting from desktop applications to web‑based tools. We analyze the characteristics of existing frameworks and, based on this analysis, we derive a set of design principles that imply low‑effort generation, extensive customization possibilities, and straightforward deployment of the resulting editors. On these grounds, we introduce \emph{\emfularp}, a purely web‑based framework for managing EMF models without any backend. The accompanying \emfularp generator maps a given Ecore model (an EMF metamodel) to a ready‑to‑use and ready-to-customize graphical editor. \emfularp editors provide `EMF consistency', that is, they not only support standard modeling operations such as creation, inspection, navigation, editing, and undo/redo, but they also handle containment and inverse references in close alignment with EMF; they also provide interoperability with existing EMF tooling through compatible de‑/serialization. A generated editor is an Angular project with designated extension points, which allows developers to customize and extend all aspects of the editor using the expressive power of Angular and its ecosystem, guided by the extension points of \emfularp. We evaluate \emfularp in terms of editor adequacy (available editing capabilities), adaptability (customization mechanisms and required effort), and robustness of the generation.
\end{abstract}


\maketitle

\section{Introduction}
\label{S:intro}

Modeling tools are essential for the construction, analysis, and maintenance of 
domain-specific languages. While the Eclipse Modeling Framework (EMF) remains 
the de-facto standard for meta-modeling and model management, there are many approaches for web-based graphical model editors  
that differ substantially in their architectural assumptions, their reliance on server-side components, and the extent to which they preserve EMF’s structural semantics. 
As a result, developers face a fragmented landscape in which no single solution combines EMF compatibility, client-side execution, and customizability within mainstream web development ecosystems.

This paper introduces \emfularp, a client-side, Angular-based framework for building 
EMF-compatible web editors, together with the \emfgenerator, a generator that produces fully functional editors from Ecore meta-models. \emfularp aims to preserve EMF’s structural semantics in the browser, support expressive SVG-based visualizations, and remain extensible through idiomatic web-development practices. To evaluate the approach, we structure the paper around four research questions:

\begin{itemize}
    \researchquestion{RQ:frameworks}{Landscape Analysis:}
    What architectural characteristics and EMF-\-re\-lated capabilities do existing web‑based modeling frameworks provide? In this manner, we inform and justify the design principles and architecture of \emfularp.

    \researchquestion{RQ:emfularp}{Framework Adequacy:}
To what extent does the \emfularp framework satisfy the architectural and EMF‑related criteria identified in the framework analysis?

  \researchquestion{RQ:generation}{Generation Robustness:}
  Does the \emfgenerator generate working web editors for real-world Ecore models without special tailoring? Can the \emfgenerator handle a diverse set of metamodels with varying size, depth, reference structures, and inheritance patterns?

  \researchquestion{RQ:customization}{Editor Adaptability:}
  Do \emfularp-based editors support customization across the established dimensions of model editor customization, and what is the effort required to realize such customizations in practice?
\end{itemize}

Each of the following sections addresses one of these research question in turn: 
Section~\ref{S:related} answers \ref{RQ:frameworks} by studying related work, mainly by means of a `landscape analysis', i.e., a criteria-based comparison of existing frameworks; Section~\ref{S:emfularp} answers \ref{RQ:emfularp} by analyzing \emfularp’s developer-facing API and architectural principles; Section~\ref{S:generation} answers \ref{RQ:generation} through a robustness validation using real-world metamodels; and Section~\ref{S:customization} answers \ref{RQ:customization} by assessing and demonstrating \emfularp’ customization mechanisms~---~relative to general customization requirements, which we identify. Together, these results provide a comprehensive evaluation of \emfularp as a client-side, EMF-compatible, and customizable foundation for web-based modeling tools, as we conclude in Section~\ref{S:concl}.

\section{Related Work Incl.\ Landscape Analysis}
\label{S:related}

We begin by discussing work relevant to the development of EMF-compatible graphical editors; we cover both conceptual foundations and empirical reports. Afterwards, we look at three current frameworks—Sirius, GLSP, and Gentleman in terms of their contrasting designs. On these grounds, we motivate a set of criteria for framework comparison. Ultimately, we perform a formal comparison of web‑based graphical editor frameworks.

\subsection{Conceptual and Empirical Background}

Our analysis of frameworks for the generation of EMF-compatible web-based graphical editors complements other surveys and comparisons on approaches that can help with the creation of graphical editors.
Often, these frameworks are complete language workbenches (LWBs), that unite meta-modeling with the generation of the tool for the language.
There are many studies that compare such LWBs~\cite{Erdweg13, Erdweg15, IungCMR20} with respect to the features of the developed languages. Criteria for evaluating the functionality of LWBs have also been proposed~\cite{Barash21}.

Currently, many LWBs follow a Low Code approach. Its position towards MDE is studied in detail by DiRuscio et al.~\cite{DiRuscio22}.
Popov et al.~\cite{PopovLV24} compare the extendability of Low Code development platforms.

A stronger focus on frameworks for graphical modeling exists on UX-based taxonomies like that of DeCarlo et al.~\cite{DeCaroLB22a}, which they used for developing an GLSP enhancement~\cite{DeCarloLB22b}, or the classification of graph-like modeling languages with respect to layouting by Wrobel and Scheffler~\cite{WrobelS23}.

In addition to analyzing frameworks, it is instructive to examine reports from developers who have built concrete graphical editors using these technologies.
Most helpful are comparisons, like that of Sirius and Graphiti for the development of a graphical editor for EMF models~\cite{ViyovicMP14}.
Recently, Ali et al.~\cite{Ali24} described the development of an EMF‑based modeling tool at Rolls‑Royce and compared their experiences with Sirius and GLSP.
They reported openly on success factors and obstacles that made them prefer GLSP over Sirius.
A tool to mine user interactions in Sirius- and GLSP- based model editing tools~\cite{DehghaniBW23} enables comparison in terms of editor UX.
Kolovos et al.~\cite{Kolovos15} study the usage of EMF technologies on open source frameworks, assessing its rich adoption.

\subsection{Framework Architectures: Sirius, GLSP, Gentleman}

To understand the architectural landscape of EMF‑compatible graphical editors, we examine three frameworks: Sirius and GLSP as the two established and widely used approaches for EMF‑based graphical modeling, and Gentleman, which represents the only fully client‑side, serverless architecture in the current landscape. Together, these three frameworks illustrate the main architectural options that motivate the criteria introduced in the next subsection.

\paragraph{Sirius and SiriusWeb}
Sirius~\cite{sirius, MadiotP15} is EMF's classical framework for generating graphical editors for EMF models.
As a low‑code framework, its graphical representations are defined largely through configuration rather than code. This approach simplifies editor creation but also constrains expressiveness: although Sirius uses SVG as its graphical basis, it provides only a small set of predefined diagram types (such as a generic graph layout, a sequence diagram, or a Gantt chart). Mixing different layout paradigms within a single representation is not supported.

Despite these limitations, Sirius has a long tradition in the Eclipse ecosystem and is widely used in teaching and industrial MDE settings~\cite{MaschottaJZ16, FriedlALV20, Shaked22, Antanasijević23}. Over the years, it has been extended with additional generation mechanisms~\cite{BediniMZ21} and support for annotated meta‑models~\cite{RichardsonKG24}.

SiriusWeb~\cite{GiraudetBBCD24} ports the Sirius concepts to a modern web technology stack in a server‑centric architecture: a SpringBoot‑based Java backend maintains the EMF model and evaluates AQL expressions, while a React frontend renders the graphical view. This design keeps most editing logic on the server, making serverless deployment impossible.

\paragraph{GLSP}
The Graphical Language Server Platform (GLSP)~\cite{BorkLO23, BorkL23} takes a fundamentally different approach from low‑code frameworks like Sirius. Rather than providing predefined diagram types or configuration‑based abstractions, GLSP offers a protocol‑centric architecture using the Language Server Protocol (LSP): a server maintains and manipulates the model, while the client renders a graphical view and forwards user commands.

This architecture is particularly attractive when external model management is required or when multiple compatible editors (e.g., a desktop and a web version) must be supported. Features such as undo/redo and the rendering of validation problems come naturally with the LSP-based design. GLSP is actively developed, but its complexity and the limited availability of up-to-date documentation and examples have been noted in recent evaluations~\cite{Ali24, MetinB25}.

GLSP offers two server implementations: one in Java and one in NodeJS. The Java server can operate directly on EMF models, whereas the NodeJS server requires developers to provide a TypeScript-based JSON model implementation, thereby moving responsibility for EMF-level consistency guarantees to the developer. Due to its higher resource requirements, the Java server is often discouraged in practice, although it remains part of the official starter templates~\cite{glsp}.

On the client side, templates exist for Eclipse Theia, Eclipse RCP, and VS Code, as well as a pure TypeScript client. While some integration notes suggest that both server and client could theoretically run in a browser environment~\cite{glsp, MetinB23}, no concrete examples exist, and the NodeJS-based server is fundamentally designed for server-side execution.

\paragraph{Gentleman}

Gentleman is, to our knowledge, the only framework in the current landscape that achieves a fully client‑side, serverless execution model. It is implemented entirely in JavaScript and can be embedded into any JavaScript- or TypeScript-based web page.
Architecturally, Gentleman is a projectional language workbench: instead of providing a dedicated graphical modeling canvas, it renders editors using HTML elements and CSS. This makes it suitable for lightweight projectional editors but limits its applicability for detailed graphical modeling. Gentleman offers an experimental mapping from Ecore to its internal meta‑model, but the mapping covers only basic concepts and does not preserve EMF’s structural semantics; the resulting models cannot be imported back into EMF. Gentleman therefore targets a different class of modeling tools and serves primarily as a reference point for browser‑only deployment rather than EMF interoperability.

\subsection{Criteria-based Evaluation}
\label{S:frameworks}

The analysis in the previous subsection highlights recurring architectural and technological themes across existing web‑based graphical modeling frameworks. Based on these observations, we derive the following criteria for EMF-compatible, extensible client-side graphical modeling frameworks:

\begin{itemize}
\item \textbf{Ecore Compatibility:}  
      Whether a framework can work directly with an existing Ecore meta-model in a way that preserves EMF’s meta-level semantics (e.g., containment, opposites, multiplicities), either natively or through a conversion step.  
      \emph{Why it matters:} Translating an Ecore meta-model into another meta-language introduces cognitive overhead and risks semantic drift, making it harder to maintain alignment with EMF-based tooling and developer expectations.

  \item \textbf{Model-level EMF Compatibility:}  
        Whether the concrete models produced by the editor can be serialized and deserialized in EMF-native formats.  
        \emph{Why it matters:} EMF-native persistence is required to integrate with existing toolchains.

\item \textbf{Developer Constraints:}  
    Architectural or tooling restrictions that shape how developers must implement, extend, or integrate the editor, such as protocol layers (e.g., LSP), mandatory view‑model mappings, or low‑code abstractions that limit direct control. 
      \emph{Why it matters:} Such constraints restrict how freely developers can implement custom behavior, integrate existing code, or shape the editor’s interaction and update mechanisms.

  \item \textbf{Execution and Deployment Model:}  
        Whether the framework runs purely client-side or requires a server, and what infrastructure is needed for hosting.  
        \emph{Why it matters:} Server-based frameworks require continuous compute resources and operational infrastructure, whereas client-only deployments can be hosted as static assets, which are widely supported by low-cost or free hosting services.

\item \textbf{Technology Stack:}  
      The client- and server-side technologies developers must adopt, such as specific web frameworks, languages, or build systems.  
      \emph{Why it matters:} These requirements determine the skills developers must have and the effort needed to integrate the framework into existing architectures and workflows.

\item \textbf{Graphical Basis:}  
      The rendering approach a framework uses for diagrams, and whether developers draw directly in the browser or through an intermediate protocol or view model.  
      \emph{Why it matters:} The rendering foundation determines how much control developers have over visuals and interaction, and how deeply the editor can be customized.
\end{itemize}

We now apply these criteria to the set of frameworks discussed in the recent informal comparison by Metin and Bork~\cite{MetinB25}, who contrasted their GLSP framework with several other web‑based graphical modeling tools, namely AToMPM~\cite{Syriani13}, WebGME~\cite{Maroti14}, CINCO Cloud~\cite{NaujokatLKS18}, KIELER~\cite{JaegerMJWZ16}, jjodel~\cite{Jjodel23, BRVP25}, Gentleman~\cite{DucoinS22,LafontantS25}, Dandelion~\cite{MartinezLasacaDGL23}).
We omit PictoWeb~\cite{Picto22} because it provides model visualization but no editing capabilities and add GLSP in its two server variants, Java and NodeJS, since these differ fundamentally in their meta‑model integration and EMF compatibility. The resulting comparison is shown in Table~\ref{T:comparison}.

\begin{table*}[t!]
    \centering
    \begin{small}
    \begin{tabular}{c | c c c c c c}
    Tool 
    & \makecell{Meta-Model\\with EMF Semantics} 
    & \makecell{EMF-native\\Read/Write} 
    & \makecell{Developer\\Constraints} 
    & \makecell{Client\\Tech} 
    & \makecell{Server\\Tech} 
    & \makecell{Graphical\\Basis} \\
    \hline\hline

    GLSP (Java) 
    & \makecell{EMF\\full EMF semantics}
    & EMF (XMI)
    & LSP
    & \makecell{TS (pure or NodeJS)\\or Eclipse Theia/RCP\\or VSCode} 
    & Java 
    & \makecell{SVG\\(Sprotty)} \\ 
    \hline

    GLSP (Node) 
    & \makecell{custom JSON\\no EMF semantics}
    & no EMF
    & \makecell{LSP\\custom view-model}
    & \makecell{TS (pure or NodeJS)\\or Eclipse Theia/RCP\\or VSCode} 
    & NodeJS 
    & \makecell{SVG\\(Sprotty)} \\ 
    \hline

    Sirius Web 
    & \makecell{EMF\\full semantics}
    & EMF (XMI)
    & low-code
    & React 
    & \makecell{SpringBoot\\(Java)} 
    & SVG \\ 
    \hline

    AToMPM 
    & \makecell{simplified UML\\partial semantics}
    & no EMF
    & low-code
    & NodeJS 
    & \makecell{NodeJS\\and Python} 
    & SVG \\ 
    \hline

    WebGME 
    & \makecell{GME meta-model\\partial semantics}
    & no EMF
    & low-code
    & \makecell{JS (pure or NodeJS)} 
    & NodeJS 
    & \makecell{SVG\\(Raphael)} \\ 
    \hline

    CINCO Cloud 
    & \makecell{EMF\\full semantics}
    & EMF (XMI)
    & \makecell{LSP\\low-code}
    & \makecell{Eclipse Theia\\or VSCode\\via Angular} 
    & Java 
    & \makecell{SVG\\(Sprotty)} \\ 
    \hline

    KIELER 
    & \makecell{S (none) or EMF\\(full semantics)}
    & EMF (XMI)
    & LSP
    & \makecell{NodeJS\\or VSCode} 
    & Java 
    & \makecell{SVG\\(Sprotty)} \\ 
    \hline

    jjodel 
    & \makecell{JOM\\no EMF semantics}
    & EMF planned
    & low-code
    & React 
    & NodeJS 
    & \makecell{HTML\\(JSX)} \\
    \hline

    Gentleman 
    & \makecell{UML-like\\partial semantics}
    & no EMF
    & low-code
    & JS 
    & -- 
    & HTML \\ 
    \hline

    Dandelion 
    & \makecell{graph-based (RDF)\\no EMF semantics}
    & EMF import (lossy)
    & low-code
    & React 
    & NodeJS 
    & ? \\
    \hline

    \end{tabular}
    \end{small}
    \caption{Comparison of Web-based Modeling Frameworks along the defined criteria.}
    \label{T:comparison}
\end{table*}

\subsection{Answer to \ref{RQ:frameworks}: Landscape Analysis}

Overall, the comparison shows that web-based modeling frameworks differ
substantially in their architectural assumptions, deployment requirements, and
support for EMF-based workflows.
None of the established frameworks combine full EMF compatibility — meaning
generation from an EMF meta-model, preservation of structural semantics, and
import and export of models for use in existing EMF workflows — with a
cost-efficient serverless deployment model and the expressive power of an
SVG-based web editor. 
These findings indicate a need for an approach that preserves EMF’s structural semantics, operates entirely client‑side, and remains fully customizable within mainstream web development ecosystems.

As a consequence of the landscape analysis, the following design principles emerge for frameworks aiming to provide a fully client‑side, EMF‑compatible graphical web editor:
\begin{itemize}
  \item \textbf{Preserve EMF Semantics at the Meta-Model Level:}  
        The editor should use a fully EMF-compatible internal meta-model, ensuring that EMF’s structural semantics remain intact.

  \item \textbf{Support EMF-Native Persistence:}  
        Concrete models should be readable and writable in EMF-native formats to integrate with existing EMF toolchains.

  \item \textbf{Avoid Mandatory Architectural Constraints:}  
        Frameworks should not impose protocol layers (e.g., LSP), custom view-model mappings, or low-code abstractions that restrict developer control.

  \item \textbf{Enable Fully Client-Side Execution:}  
        A modeling framework should support serverless deployment as static web assets without requiring a backend runtime.

  \item \textbf{Use a Standard Web Technology Stack:}  
        The framework should integrate naturally with mainstream web development ecosystems without requiring specialized runtimes or IDE platforms.

  \item \textbf{Provide a Direct, Standards-Compliant SVG Rendering Layer:}  
      The graphical basis should be expressive, stan\-dard‑compliant, and directly programmable, without intermediate protocols or hidden rendering layers.
      SVG is preferable over HTML‑based rendering because it offers richer graphical expressiveness and native support for shapes and transformations.

\end{itemize}

\section{The \emfularp Framework}
\label{S:emfularp}

We shortly illustrate the extension points and architectural constraints of the general-purpose web development framework Angular onto which \emfularp is built.
We then position \emfularp within an established reference architecture for web-based modeling tools. This allows us to assess \emfularp not in isolation, but in relation to a widely accepted decomposition of responsibilities across core model management, diagramming, and tool-level functionality.

Once this architectural context is established, we show how \emfularp implements these layers through a set of lightweight, client-side packages that conform to Angular’s idioms and together form a coherent and extensible foundation for
EMF-compatible web editors.

\subsection{Angular Concepts for Extensible Editors}
\label{S:angular}

Angular is a mature open-source web application framework developed by Google. It provides a component-based architecture in which applications are composed of \emph{services} and \emph{components} implemented in TypeScript.

\emph{Services} encapsulate shared logic or state, such as model access, command execution, or file management. They expose typed methods or observable event streams and are injected into other parts of the application through Angular’s dependency-injection system. Observables and subscriptions allow services and components to exchange state and notify each other of changes without relying on global variables or tightly coupled interfaces.

\emph{Components} build on services by adding a visual representation. A component
consists of a TypeScript class, an optional stylesheet, and a template that defines its rendered structure. Components communicate through typed \emph{inputs} and \emph{outputs}: inputs supply data or context, while outputs emit events that surrounding components or services can subscribe to. This enables reusable UI elements whose behavior can be customized by their embedding context.

Component templates are typically written in HTML, but Angular also supports SVG templates, allowing components to render graphical content directly inside an \texttt{<svg>} element. This makes Angular suitable for diagrammatic editors in which visual elements are represented as SVG groups.

Angular further supports \emph{content projection}, where a component declares slots into which other components or templates can be inserted. This mechanism applies equally to HTML and SVG templates and enables complex visual structures to be composed from smaller, reusable building blocks.

Together, these concepts—services, components with HTML or SVG templates, input/output communication and subscriptions, content projection, and dependency injection—form the technical foundation on which \emfularp’s extensibility mechanisms are built.

\subsection{Reference Architecture for Web-Based Modeling Tools}
Metin and Bork~\cite{MetinB25} propose a reference architecture that structures LSP- and web-based modeling tools into three layers: \emph{Core}, \emph{Diagram}, and \emph{Tool} on top of the \emph{LSP}. Each layer groups a distinct set of responsibilities and APIs, and together they define the conceptual building blocks required for interactive, model-driven editors.

While the reference architecture is presented in the context of LSP-based systems, its separation into Core, Diagram, and Tool layers generalizes naturally to modeling tools that do not rely on an LSP backend, in which case the responsibilities typically delegated to the language server — such as model-based CRUD operations and consistency management—must be implemented directly within the client-side editor. These concerns are usually handled in the Core and Tool layers, which assume a broader role in maintaining model state and ensuring consistency. As a result, the architecture remains applicable beyond LSP-based systems, with a shifted distribution of responsibilities:

\begin{itemize}
  \item \textbf{Core} provides the fundamental model management capabilities. This includes model state, containment and reference semantics, command handling, and the APIs through which higher layers create, modify, and query model elements.

  \item \textbf{Diagram} offers the visualization and interaction facilities required to present models graphically. This layer typically includes shape definitions, layout, selection, and the mapping between model elements and their visual representations.

  \item \textbf{Tool} comprises the user-facing editor functionality such as palettes, inspectors, creation logic, and integration with surrounding application frameworks. It builds on the services of the Core and Diagram layers to provide a complete editing experience.
\end{itemize}

We use this architecture as the conceptual frame for describing the structure of \emfularp.

\begin{figure*}[ht!]
\centering
    \makebox[0pt]{%
    \includegraphics[width=.99\textwidth]{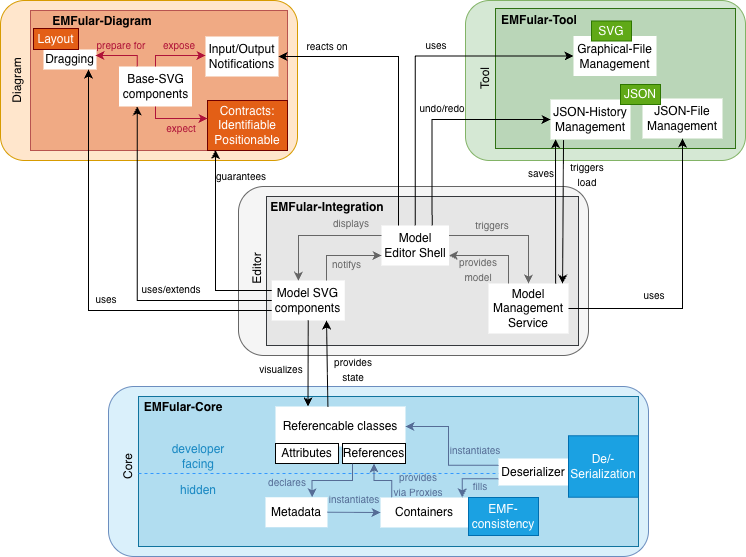}
    }
\caption{Megamodel of \emfularp}
\label{F:emfular:architecture}
\end{figure*}

\subsection{Architecture of \emfularp}

\emfularp consists of three independent libraries, \emfularc, \emfsvg, and \emfhelpers: each of them addresses one layer of the reference architecture.
They fulfill the respective responsibilities without coupling, basically by a small set of assumptions that using projects must fulfill.
\emfintegration relies on these assumptions to assemble the layer-specific libraries into ready‑to‑use editor components that can be embedded directly into an application or extended through Angular’s component and service mechanisms.

All libraries are published as open‑source packages and distributed via the JS standard package manager \texttt{npm}, enabling modular adoption and integration into existing applications.
Figure~\ref{F:emfular:architecture} summarizes the architecture of \emfularp in a megamodel~\cite{FavreLV12}.

\paragraph{\emfsvg: SVG-based Components}
\emfsvg encourages that an editor renders its graphical content on an SVG canvas and that this canvas is populated with small, reusable Angular components whose templates are written in SVG.
To support such components, the library provides explicit notions of graphical identity and position, together with a lightweight notification mechanism that emits movement events whenever an element changes its position. These assumptions enable a consistent interaction model in which connectors and arrows can attach to elements by their graphical ID and update automatically when those elements move. Building on these contracts, \emfsvg contributes reusable SVG components, a two‑layer dragging mechanism, and adaptive connectors that subscribe to movement notifications, forming a minimal but expressive foundation for constructing SVG-based diagrams in Angular.

\paragraph{\emfhelpers: Generic Modeling Utilities}
\emfhelpers assumes that the final editor operates on a canvas — HTML or SVG — that should be exportable as SVG, PNG, and JPEG, and that it maintains some JSON‑serializable model whose structure is not known in advance. Based on these assumptions, the library provides generic file I/O utilities for loading and saving text and JSON files, exporting canvas content, and integrating these operations with the browser’s download facilities. A lightweight history mechanism stores arbitrary JSON snapshots in a circular buffer and persists them in \texttt{localStorage}, enabling undo/redo and session recovery without making any assumptions about the meaning or schema of the stored data.
These utilities abstract the tool‑level functionality commonly required in web‑based editors into idiomatic Angular services and components, while remaining entirely independent of EMF semantics or diagramming concerns.

\paragraph{\emfularc: Runtime and Persistence}
\emfularc is a pure TypeScript library that provides EMF‑style model semantics without depending on Angular or any UI framework, making it usable in other TypeScript‑based environments such as Node.js or React. Its purpose is to expose models through a developer‑facing API that behaves like ordinary TypeScript objects with explicit attributes and references, while internally guaranteeing the strong consistency constraints of EMF.

At the developer level, model classes declare their structure explicitly: attributes are ordinary TypeScript fields, and references appear as single‑valued or list‑valued
properties of the target type. Behind this familiar API, however, each reference is backed by a lightweight proxy that forwards all mutating operations to a hidden layer of containers, selected according to EMF‑style metadata attached to the
reference declaration.
These containers implement the semantics of EMF references — containment, bidirectional opposites, deletion cascades — and ensure that all modifications preserve the invariants of a well‑formed EMF object graph.
Convenience methods for reordering lists and deleting elements recursively complement the developer‑facing API.
In this way, the declarative structure of the model and the runtime semantics that enforce EMF‑style consistency remain cleanly separated yet tightly connected.

For persistence, \emfularc adopts EMF‑Jackson, EMF’s native JSON‑based serialization format, ensuring direct interoperability with existing EMF workflows.
Its JSON structure is web‑native and permissive, allowing applications to persist additional information alongside the structural model without breaking compatibility.
Serialization is driven directly by the attribute and reference declarations of a model class: any field marked with the attribute decorator is written to JSON without requiring changes to the metamodel, which makes it straightforward to persist additional information such as layout or positions.
Reference serialization is delegated to the containers, which produce nested JSON for containment references and EMF‑Jackson URI‑fragment strings for cross‑references.
Deserialization reconstructs attributes and the containment hierarchy in a first pass and resolves all cross‑references in a second, yielding a fully consistent EMF object graph.

\paragraph{\emfintegration: Ready‑to‑Use Editor Components}
\emfintegration is the only package that depends on all others. It assembles \emfularc, \emfsvg, and \emfhelpers into a set of ready‑to‑use, model‑agnostic editor components that can be embedded directly or extended through Angular’s component and service mechanisms.
Building on JSON‑serializable \emfularc models and an SVG‑\-ba\-sed canvas, its central element is an extensible editor shell that provides a file‑level bar with standard model operations (load, save, export, undo/redo), a central SVG canvas with a projection slot for model‑specific visualizations, and an optional model‑editing bar that applications can populate with domain‑specific creation or layout actions.

A stateful model‑management service supplies the editor shell with the current model instance, integrates the history mechanism for undo and redo, and exposes a simple API for creating, loading, and saving models. This service forms the operational backbone of the editor: all components interact with the same model state, and applications can refine the workflow by extending the service without modifying the editor shell itself.

Beyond the shell, \emfintegration offers a small set of ready‑to‑use SVG components and interaction services for constructing tree‑based graphical editors for EMF-models.
These components render containment hierarchies as nested SVG boxes and provide matching interaction services for selection, navigation, and assignment. While intentionally model‑agnostic, they form a practical foundation from which applications can build richer, domain‑specific visualizations.

\subsection{Answer to \ref{RQ:emfularp}: Framework Adequacy}

The \emfularp framework provides the essential capabilities expected from EMF‑based model editors. Its core data layer supports EMF‑compatible object graphs with attributes, references, containment, opposites, and deletion cascades, ensuring that editing operations preserve EMF semantics. Editors built on top of the framework can create, inspect, navigate, and modify models uniformly, and the integrated history mechanism offers undo and redo at the level of model operations. Persistence is handled through EMF‑Jackson, enabling EMF‑native JSON serialization without custom formats. The architecture exposes metadata and extension points to integrate structural checks (validation).

\emfularp imposes no mandatory protocol layers or low‑code abstractions; it is implemented entirely in standard web technologies and executes fully client‑side as static assets. Its rendering layer is based on direct, standards‑compliant SVG, which Angular’s component mechanism encapsulates into reusable building blocks. Taken together, these properties show that \emfularp meets all six criteria and is adequate for building interactive EMF‑style web editors without additional infrastructure.

\section{The \emfgenerator}
\label{S:generation}

The \emfgenerator consumes a single Ecore meta-model and produces a corresponding initial web editor as a stand‑alone Angular project that depends solely on the \emfularp packages available via \emph{npm}.
The generated project exposes all editor logic directly in Angular components and services, without additional protocols or specialized tooling, giving developers an immediately usable and fully transparent starting point for further extension.

\subsection{Generator Architecture and Workflow}

\begin{figure}[t!]
\centering
\begin{small}
\begin{tikzpicture}[
  node distance=5mm,
  box/.style={
    rectangle,
    draw,
    rounded corners,
    align=center,
    minimum width=40mm,
    minimum height=8mm
  },
  userbox/.style={
    rectangle,
    draw,
    dashed,
    rounded corners,
    align=center,
    minimum width=30mm,
    minimum height=8mm
  },
  arrow/.style={->, thick}
]

\node[box] (ecore) {Ecore File\\(single \texttt{.ecore})};

\node[box, below=of ecore] (parser) {Ecore Parser\\
  \footnotesize extracts EPackage\\
  \footnotesize produces JSON meta-model};

\node[userbox, right=8mm of parser] (user) {User selects\\root class};

\node[box, below=10mm of parser] (templates) {Template Engine\\
  \footnotesize Angular workspace templates\\
  \footnotesize domain TS classes\\
  \footnotesize editor and model service};

\node[box, below=of templates] (project) {Generated Angular Application\\
  \footnotesize uses EMFular via npm\\
  \footnotesize fully modifiable\\
  \footnotesize deterministic output};

\draw[arrow] (ecore) -- (parser);
\draw[arrow] (templates) -- (project);

\draw[arrow] (parser.east) -- (user.west);
\draw[arrow] (user.south) |- (templates.east);

\end{tikzpicture}
\end{small}
\caption{Generation pipeline of the \emfgenerator: parsing the Ecore model, selecting a root class, and instantiating templates to produce the Angular project.}
\label{F:generator-pipeline}
\end{figure}
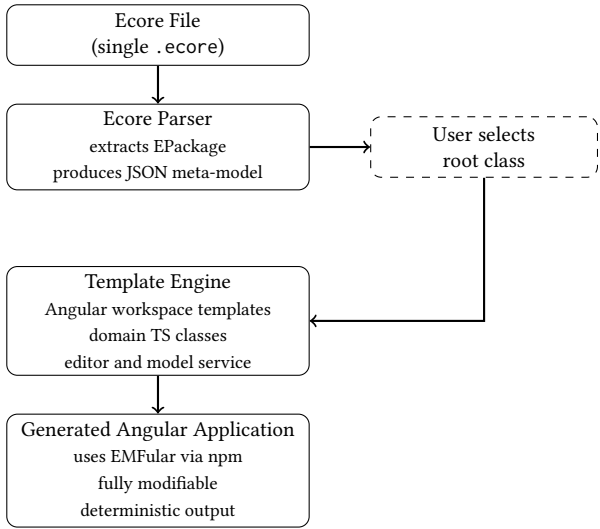

The generator follows a simple two-stage workflow, shown in Figure~\ref{F:generator-pipeline}. First, it processes the input Ecore model and extracts a uniform JSON representation that contains the structural information needed for code generation.

\subsection{Template-based Generation}

\begin{figure*}[ht!]
\centering
    \makebox[0pt]{%
    \includegraphics[width=0.99\textwidth]{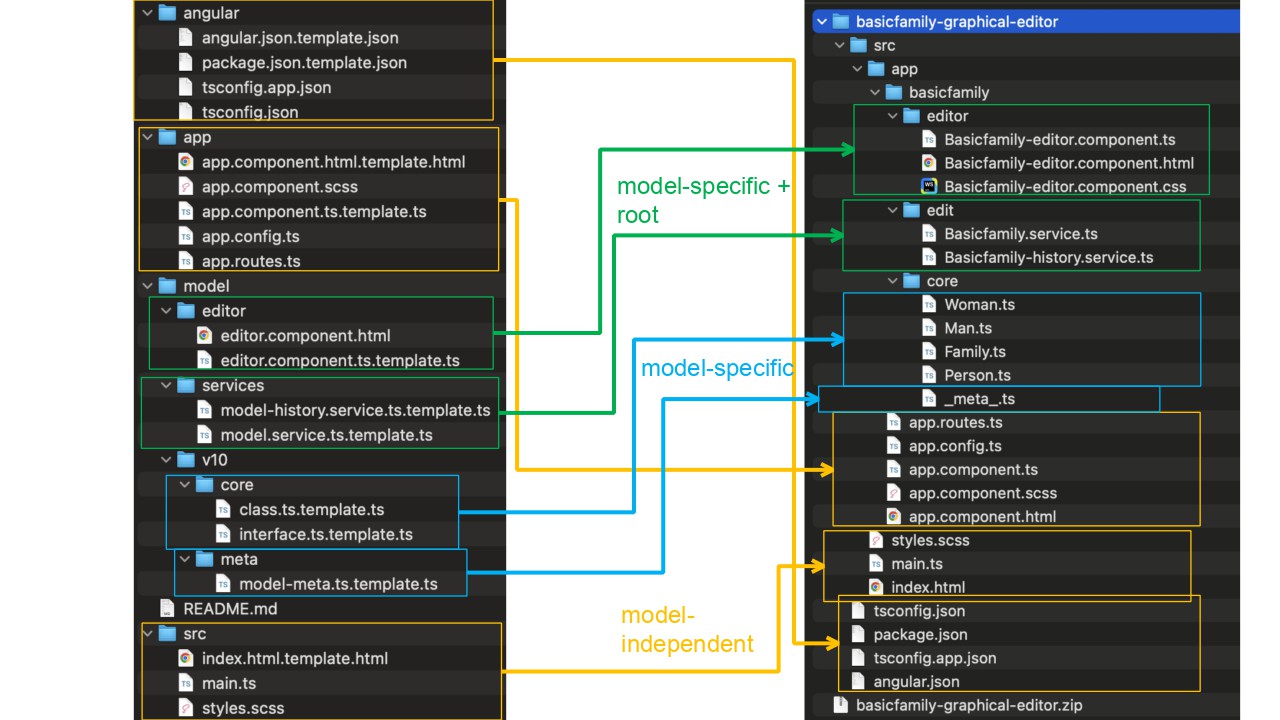}
    }
\caption{Correspondence between the generator templates and the generated project}
\label{F:generation:templates}
\end{figure*}

Before generation begins, the user is asked to select a root class from the set of applicable EClasses.
This root determines the serialization root and thus the top‑level element that the generated editor loads, displays, and persists.
In the second stage, the generator then instantiates a set of templates to assemble the Angular project.

Figure~\ref{F:generation:templates} illustrates how the generator assembles a complete Angular project from its internal template structure. The left side shows the generator’s template folders, grouped into model‑agnostic layers (yellow) and a model‑specific layer. The latter is divided into two parts: a faithful replication of the Ecore meta‑model as TypeScript classes together with a corresponding JSON meta‑model (blue), and editor‑level templates (green) that integrate this model information into the generated editor and its services.

The model‑agnostic layer (yellow) is identical for all generated editors, aside from trivial adaptations such as the project name. The model‑specific layer (blue) reflects the detailed structure of the given Ecore model and therefore scales with model size. The editor templates (green) depend on both the model and the user‑selected root class: they generate one creation command per concrete class, register palette entries, and configure the editor’s state and history management. Although larger models lead to more creation commands and palette entries, the editor layer itself always consists of the same Angular primitives: an editor component and two accompanying model‑level services.

\subsection{Validation}

Several end‑to‑end tests, using models of different size and structure, were performed. These tests consist of uploading an \texttt{.ecore} file, selecting a root class, unpacking the generated workspace, installing its dependencies via \texttt{npm install}, and launching the editor using \texttt{ng serve}. The generator handles both classical examples (such as \texttt{ecore.ecore}) and large real‑world models, including the IFC schema with nearly 700 EClasses, 7 EDataTypes and 164 EEnums. Thus, the workflow is functional and the generation- and web-based approach can handle large and nested metamodels. Exhaustive and fully automated testing with the Ecore models of Atlantic Zoo has revealed some issues such as usage of subpackages requiring more refined treatment and name clashes between metamodel elements and \emfularp{} requiring resolution. These issues are being addressed.

\subsection{Answer to \ref{RQ:generation}: Generator Robustness}

Our experiments indicate that \emfgenerator can produce working web editors for real‑world Ecore models without model-specific tailoring. The generator behaves deterministically—identical input models and root selections yield identical projects—and the architecture scales to metamodels with substantial size, depth, reference structures, and inheritance hierarchies. Although development is ongoing and some issues were identified during testing, the generator already handles both classical EMF examples and large, nested metamodels such as the IFC schema.

\section{Customization}
\label{S:customization}

Beyond generating a functional baseline editor, practical modeling tools must support adaptation to domain-specific notations, workflows, and integration needs. Customization is therefore a central aspect of evaluating EMFular's capabilities: it determines whether the generated editors can be meaningfully extended, and whether such extensions remain maintainable.

In this section, we first identify general customization dimensions for graphical model editors, then describe how EMFular realizes these capabilities, and finally demonstrate the customization process using a concrete example.

\subsection{Customization Requirements for Graphical Model Editors}

Graphical modeling editors must accommodate a wide range of domain- and workflow-specific needs.
Across existing low-code modeling frameworks like Sirius or AToMPM, several recurring customization dimensions have emerged that shape both the expressiveness of the resulting editors and the effort required to adapt them. 
These dimensions form a general baseline of expectations against which any modeling framework or generator must be evaluated.

\begin{itemize}
  \item \textbf{General look and feel.} 
  Modeling tools are often embedded into larger ecosystems and therefore require control over visual appearance, including color schemes, typography, layout structure, and branding. 
  Such adaptations ensure visual coherence with existing tools and support usability requirements specific to the target domain.

  \item \textbf{Graphical representation of model elements.} 
  A core requirement is fine-grained control over how model concepts are visualized: which elements appear as nodes or edges, their styling, how they are grouped or layered, and which layout strategies govern their spatial arrangement. 
  These decisions strongly influence the readability and semantic adequacy of the resulting diagrams.

  \item \textbf{Domain-specific perspectives and workflows.} 
  Many modeling scenarios rely on multiple (complementary) views of the same underlying model. 
  These perspectives emphasize different subsets of concepts, provide tailored visual models, palettes or toolbars, and guide users through workflow-specific tasks. 
  Supporting such perspectives requires extensibility for several graphical model representations, configurable interaction points, navigation structures, and editing hooks.

  \item \textbf{Integration of external libraries and services.} 
  Modern modeling environments frequently incorporate external components such as layout engines, visualization toolkits, validation frameworks, or domain-specific UI widgets. 
  Effective customization therefore requires well-defined integration points that allow such libraries to be introduced without modifying the underlying platform.
\end{itemize}

\subsection{Customization Capabilities in EMFular}
\label{SS:customization-realization}

\emfularp is built on top of Angular and therefore inherits the framework's extensibility mechanisms—component templates (HTML and SVG), typed input/output bindings, and content projection—as introduced in Section~\ref{S:angular}. 
These mechanisms form the basis on which EMFular realizes the customization dimensions identified above.

The \emfgenerator produces a complete, idiomatic Angular workspace that includes the domain model, a stateful model service, and a minimal initial editor. 
Although a ready-to-use tree component is available in the \emfintegration package, the generator deliberately reconstructs it inside the generated project. 
This replayed construction exposes the relevant extension points directly to the developer.

\begin{itemize}
  \item \textbf{General look and feel.}
  Since the generated editor is a standard Angular application, its visual appearance can be adapted using familiar mechanisms such as CSS, Angular Material theming, and template modification. 
  Developers can adjust layout structure, typography, color schemes, and branding by editing ordinary components and style sheets, ensuring seamless integration into existing web applications or corporate design systems.

  \item \textbf{Graphical representation of model elements.}
  \emfularp encourages a component-based approach to graphical modeling: each model element can be associated with a dedicated Angular component whose template defines its visual representation, and whose inputs and outputs open natural extension points to insert state and context or receive outputs on user-triggered events. 
  To give developers complete control over all aspects of a visualization, EMFular strongly recommends the use of SVG templates for these components. 
  SVG components can be composed, nested, and grouped in the same way as HTML components, but offer far richer control over shapes, layering, and spatial arrangement—essentially everything that the SVG standard can express. 
  EMFular supports this style of modeling through the \emfsvg package, which provides reusable building blocks for common diagramming tasks such as text placement, bounding boxes, and drag handles, while leaving the overall structure and notation entirely under the developer’s control.

\item \textbf{Domain-specific perspectives and workflows.}
  Perspectives in EMFular are Angular components that render the model—or the relevant subset of it—using either HTML or SVG templates. 
  HTML-based perspectives support form-oriented or inspector-style workflows, while SVG-based perspectives enable graphical modeling. The generated tree-based editor serves as an example of an SVG-based perspective: it is a standard HTML component that contains an \texttt{<svg>} element as its main modeling area and exposes a content-projection slot for inserting the root component of a domain-specific SVG-based visualization. 
  This allows the perspective to define the overall interaction frame—such as selection handling, toolbars, or navigation structures—while delegating the actual rendering of model elements to the projected component hierarchy. 
  Additional perspectives can be introduced by providing alternative components that reuse the same model service, and standard Angular interaction patterns (e.g., modal dialogs, detail views, or workflow-specific commands) can be integrated naturally since perspectives are ordinary components embedded in the surrounding application.

  \item \textbf{Integration of external libraries and services.}
  The generated workspace follows Angular’s dependency-injection conventions, making it straightforward to integrate external visualization libraries, layout engines, validation frameworks, or domain-specific UI widgets. 
  New services can be registered alongside or in place of generated ones, and the component-based architecture ensures that external functionality can be introduced without modifying the generator or the core framework libraries.
\end{itemize}

Together, these mechanisms allow EMFular-based editors to realize the full customization space identified in the previous subsection while remaining entirely within idiomatic Angular development practices.

\subsection{Example: Extending the BasicFamily Editor}
\label{SS:example-extension}

To illustrate the customization mechanisms described above, we replay two well-known Sirius tutorials of the \emph{BasicFamily} metamodel. 
Starting from the generated EMFular editor—which provides the model service, the tree-based perspective, and the surrounding application shell—we first replace the default SVG tree perspective by implementing the graphical notation from the Sirius \emph{Starter Tutorial} using domain-specific SVG components. 
We then add an HTML-based details perspective corresponding to the Sirius \emph{Compartments Tutorial}, opened via a double-click on a family member and presenting detailed information in a form-oriented view. 
The editor also retains the single-click modal interaction from the generated tree-based perspective, which is used here to change names and assign parents. 
This illustrates that workflow-specific interactions can be adapted or extended using standard Angular components and services.

\begin{figure*}[ht!]
\centering
    \makebox[0pt]{%
    \includegraphics[width=0.99\textwidth]{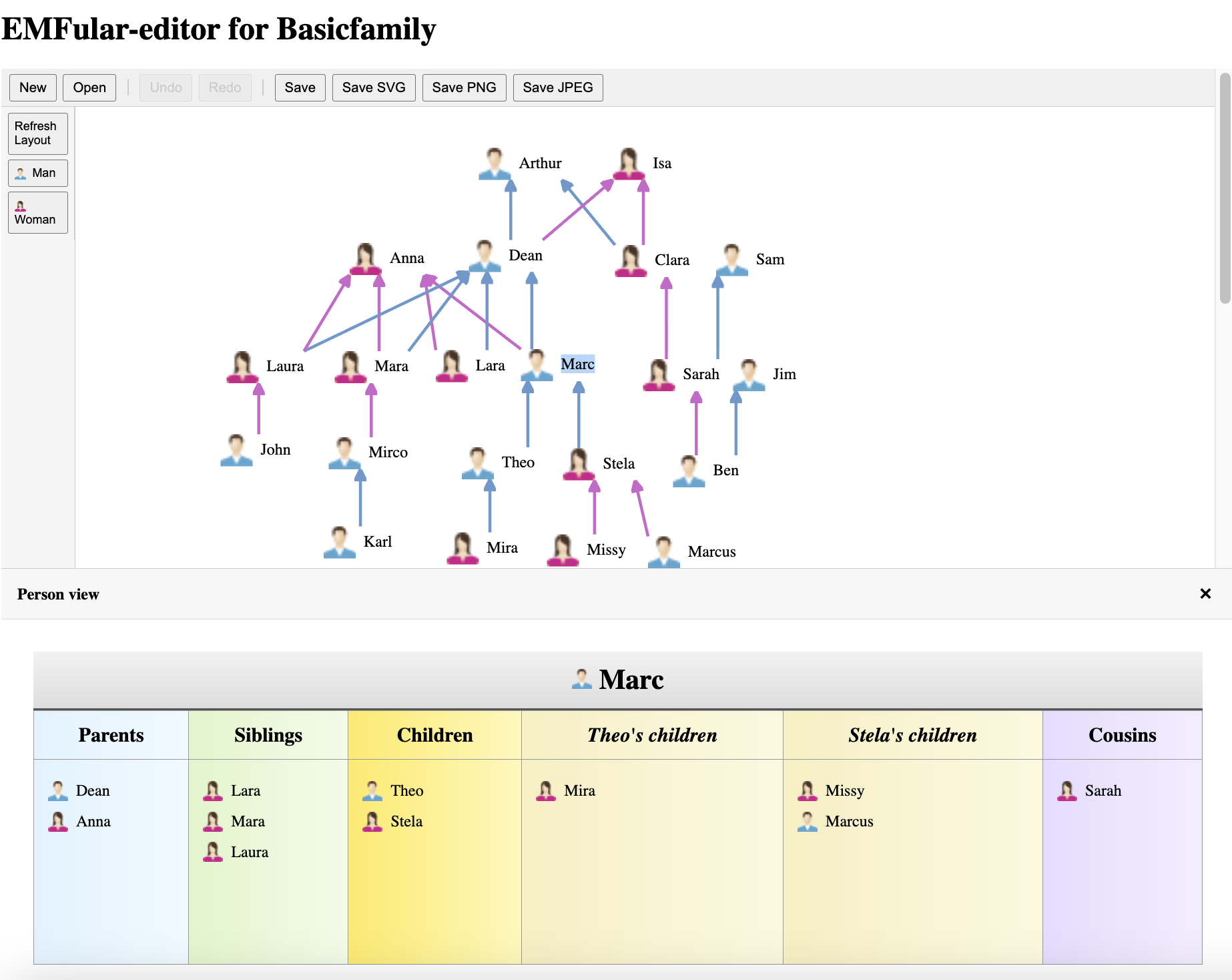}
    }
\caption{Customized BasicFamily editor with SVG-based modeling canvas (top) and HTML-based details view (bottom)}
\label{F:basicfamily}
\end{figure*}

Fig.~\ref{F:basicfamily} illustrates the resulting editor for the \emph{BasicFamily} metamodel, combining an SVG-based graphical perspective and an HTML-based details perspective. 
The complete source code of the extended editor, showing its evolvement from the generated project in small commits, is available in an anonymized repository.\footnote{\url{https://github.com/emfular-demos/basicfamily-ge}}
A live instance of the customized editor, deployed via GitHub Pages from the same repository, is also available for interactive exploration.\footnote{\url{https://emfular-demos.github.io/basicfamily-ge/}}

The required modifications to the generated editor are small and localized. They fall into a few clearly separated areas:

\begin{itemize}
  \item \textbf{Model-level adaptations.}  
  The generated \texttt{Person} class already contains stubs for its derived features, \texttt{mother} and \texttt{father}, that we implement, using the stateful reference \texttt{parents}.
  We also add a helper to set father and mother, that falls back to parents as well. 
  Two simple predicates (\texttt{isMan}, \texttt{isWoman}) are introduced to support a unified graphical component for persons.

  \item \textbf{Domain-specific SVG components.} 
The graphical notation is implemented using a single domain-specific \emph{Person} component that renders individuals, selects the appropriate icon, and visualizes father and mother relationships as colored arrows. This component is instantiated for each model element by a lightweight \emph{Family} component, which iterates over all persons. By replacing the projection of the default svg-tree component with that of the family, the whole view is adapted.

\item \textbf{Positioning and layout persistence.}
We support drag-based person positioning through EMFular’s SVG utilities for dragging and arrow routing. 
To persist the resulting layout, we store each person’s position as a model attribute rather than component state and mark it for serialization. 
This ensures that the graphical arrangement is saved with the model and restored on load, providing stable, model-level layout metadata.

As an alternative to manual positioning, we integrate the ELK layouting engine as a small Angular service. 
Invoking this service from a button on the model-specific panel computes a fresh layout for the current model and updates the stored bounding boxes accordingly. 
This demonstrates how external libraries can be incorporated with minimal effort through Angular’s dependency-injection mechanism.

\item \textbf{Interaction adaptations.}
The graphical \texttt{PersonComponent} does not prescribe any interaction behaviour; instead, it simply exposes user events (such as click and double-click) to its surrounding editor. Each perspective interprets these events independently: the SVG-based perspective opens a domain-specific modal for editing names and parent relationships on single click, while the HTML-based details view updates on double click. Because the graphical component itself remains unchanged, the same rendering logic can be reused across perspectives and editors and integrated into different workflows with minimal effort.

\end{itemize}

\subsection{Answer to \ref{RQ:customization}: Editor Adaptability}

EMFular-based editors support all four established customization dimensions through the ordinary extension mechanisms of Angular. Visual appearance and layout can be adapted by modifying the top-level components and style sheets; graphical notations are defined by domain-specific SVG components with full control over rendering and clear interaction points via exposed user events; perspectives and workflows are introduced as additional Angular components that reuse the shared model service and bind to the outputs of graphical components; and external libraries integrate cleanly through dependency injection.

The BasicFamily case study shows that such customizations require only small, localized changes to the generated project, indicating that EMFular enables expressive editor adaptations with modest and predictable development effort.

\section{Conclusion}
\label{S:concl}

\begin{sloppypar}
As our landscape analysis shows, EMF-compatible, semantics-preserving graphical editors are currently only available through heavy Java-based server infrastructures that execute an EMF runtime and impose additional development constraints through low-code abstractions or protocol layers such as LSP.
\end{sloppypar}

\emfularp offers a lightweight, purely client-side alternative for creating editors that preserve EMF semantics without requiring any backend. It provides the essential capabilities for EMF-compatible editing~---~creation, inspection, navigation, modification, and undo/redo~---~while exposing clear extension points for validation. The accompanying \emfgenerator further reduces developer effort by producing an initial tree-based editor with explicit extension points directly from an uploaded Ecore file. Editors generated in this way remain fully modifiable Angular applications and can be customized along the established dimensions of model-editor variability.

These results demonstrate that a browser-native, EMF-compatible modeling stack is both feasible and effective. Future work will focus on strengthening validation support and extending the generator to cover more complex metamodel constructs such as sub-packages, further maturing \emfularp and its generator into a robust foundation for web-based modeling tools.

Over the last 18 months, \emfularp evolved in parallel to our DSML, KEML~\cite{GöbelL24}\footnote{\url{https://keml-group.github.io/web-editor/} for editor access, repository \url{https://github.com/keml-group/keml.web}}, for which we required an EMF-based fully client-side editor.
Because EMFular-based editors run directly in the browser without any installation effort, we were able to conduct ad-hoc experiments with non‑technical students~\cite{GZDLP2026}, which would have been impractical with desktop-based tooling.
These studies not only informed the evolution of KEML itself but also provided usability insights into general editor interaction patterns that we fed back into \emfularp.

As more users apply EMFular to their own DSMLs, their feedback and recurring usage patterns will guide the next development steps and support a broader ecosystem of browser-native EMF‑compatible editors across diverse domains.

\bibliography{main}

\end{document}